\documentclass[conference]{IEEEtran}

\usepackage{cite}
\usepackage{comment}
\ifCLASSINFOpdf
  \usepackage[pdftex]{graphicx}
\else
    \usepackage[dvips]{graphicx}
\fi
\usepackage{multicol}
\usepackage{multirow}

\usepackage{hyperref}
\usepackage{color}

\usepackage{amsmath}
\usepackage{algorithm}
\usepackage{algpseudocode}
\usepackage{listings}
\usepackage{xcolor}
\usepackage{amssymb}

\usepackage{pifont}
\usepackage[dvipsnames]{xcolor}

\usepackage{tcolorbox}
\usepackage{rotating}
\usepackage{textpos}

\definecolor{codegreen}{rgb}{0,0.6,0}
\definecolor{codegray}{rgb}{0.5,0.5,0.5}
\definecolor{codepurple}{rgb}{0.58,0,0.82}
\definecolor{backcolour}{rgb}{0.95,0.95,0.92}
\lstdefinestyle{cppstyle}{
    language=C++,
    backgroundcolor=\color{backcolour},   
    commentstyle=\color{codegreen},
    keywordstyle=\color{magenta},
    numberstyle=\tiny\color{codegray},
    stringstyle=\color{codepurple},
    basicstyle=\ttfamily\scriptsize, 
    numbers=left,
    breakatwhitespace=false,         
    breaklines=true,                 
    captionpos=b,                    
    keepspaces=true,                 
    numbersep=5pt,                  
    showspaces=false,                
    showstringspaces=false,
    showtabs=false,                  
    tabsize=2
}

\definecolor{highlight1}{RGB}{255, 204, 204}  
\definecolor{highlight2}{RGB}{255, 229, 204}  
\definecolor{highlight3}{RGB}{255, 255, 204}  
\definecolor{highlight4}{RGB}{229, 255, 204}  
\definecolor{highlight5}{RGB}{204, 255, 229}  
\definecolor{highlight6}{RGB}{204, 255, 255}  
\definecolor{highlight7}{RGB}{204, 229, 255}  
\definecolor{highlight8}{RGB}{229, 204, 255}  
\definecolor{highlight9}{RGB}{255, 204, 255}  
\definecolor{highlight10}{RGB}{192, 192, 192} 

\usepackage{booktabs}
\usepackage{array}

\ifCLASSOPTIONcompsoc
  \usepackage[caption=false,font=normalsize,labelfont=sf,textfont=sf]{subfig}
\else
  \usepackage[caption=false,font=footnotesize]{subfig}
\fi

\usepackage{url}

\usepackage{textpos}

\begin{document}
\newcounter{obsno}
\definecolor{textHighlight}{RGB}{227,242,253}
\newcommand{\observation}[1]{
    \begin{tcolorbox}[width=\linewidth, colback=textHighlight,left=2pt,right=2pt,top=2pt,bottom=2pt]
        \textbf{{\small \MakeUppercase{Observation} }\refstepcounter{obsno}\Roman{obsno}:} {\small #1}
    \end{tcolorbox}
}

\title{Making Room for AI: Multi-GPU Molecular Dynamics with Deep Potentials in GROMACS}

\author{
\IEEEauthorblockN{
 Luca~Pennati\textsuperscript{*}, Andong~Hu\textsuperscript{*}, Ivy~Peng\textsuperscript{*},
 Lukas~Müllender\textsuperscript{*},
 Stefano~Markidis\textsuperscript{*}}
 \IEEEauthorblockA{\textit{\textsuperscript{*}KTH Royal Institute of Technology},
 Stockholm, Sweden \\ \{pennati,andonghu,bopeng,markidis\}@kth.se \\
 lukas.mullender@scilifelab.se }
}


\maketitle

\begin{abstract}
GROMACS is a de-facto standard for classical Molecular Dynamics (MD). The rise of AI-driven interatomic potentials that pursue near-quantum accuracy at MD throughput now poses a significant challenge: embedding neural-network inference into multi-GPU simulations retaining high-performance. In this work, we integrate the MLIP framework DeePMD-kit into GROMACS, enabling domain-decomposed, GPU-accelerated inference across multi-node systems. We extend the GROMACS \texttt{NNPot} interface with a DeePMD backend, and we introduce a domain decomposition layer decoupled from the main simulation. The inference is executed concurrently on all processes, with two MPI collectives used each step to broadcast coordinates and to aggregate and redistribute forces.
We train an in-house \mbox{DPA-1 model} (1.6 M parameters) on a dataset of solvated protein fragments. We validate the implementation on a small protein system, then we benchmark the GROMACS-DeePMD integration with a 15{,}668 atom protein on NVIDIA A100 and AMD MI250x GPUs up to 32 devices.
Strong-scaling efficiency reaches 66\% at 16 devices and 40\% at 32; weak-scaling efficiency is $\sim $80\% to 16 devices and reaches 48\% (MI250x) and 40\% (A100) at 32 devices.
Profiling with the \texttt{ROCm System profiler} shows that $>$90\% of the wall time is spent in DeePMD inference, while MPI collectives contribute $\lesssim$10\%, primarily since they act as a global synchronization point. The principal bottlenecks are the irreducible ghost-atom cost set by the cutoff radius, confirmed by a simple throughput model, and load imbalance across ranks. 
These results demonstrate that production MD with near \emph{ab initio} fidelity is feasible at scale in GROMACS.
\end{abstract}


\IEEEpeerreviewmaketitle

\section{Introduction}
GROMACS~\cite{pronk2013_gromacs,abraham2015gromacs} has grown from its early-1990s origins into a de-facto standard open-source Molecular Dynamics~(MD) engine for biomolecular and materials simulations. Its focus on performance and performance portability, from desktops to leadership-class supercomputers, has led to a broad adoption across the MD simulation community~\cite{pall2020gromacs}. However, the MD  simulation landscape is now shifting. Machine-learning interatomic potentials trained on quantum-mechanical reference data promise near-\emph{ab initio} accuracy. Bringing these AI workloads into GROMACS poses a new challenge, including coordinating neural-network inference with domain decomposition, neighbor lists, and particle-mesh calculations across multiple GPUs and nodes. This work addresses this challenge by integrating AI into GROMACS, enabling multi-GPU MD with machine-learning potentials.
\begin{figure}[t]
    \centering
    \includegraphics[width=0.99\linewidth]{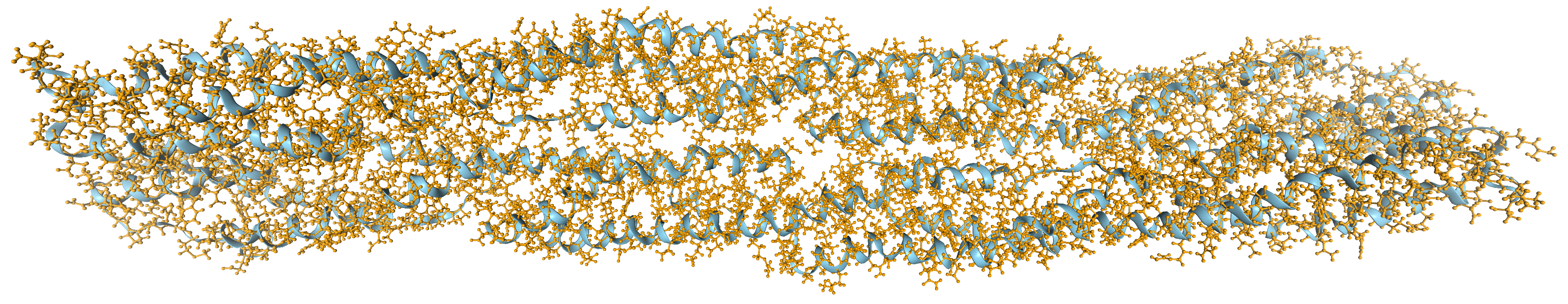}
    \caption{Structure of the 1HCI protein (15,668 atoms) obtained after 500 time steps of a multi-GPU domain decomposed \mbox{GROMACS-DeePMD} simulation, using an in-house DPA-1 model.}
    \label{fig_1HCI_structure}
\end{figure}

GROMACS is a classical MD engine, where interactions are evaluated using empirical force fields, such as AMBER~\cite{Cornell1995_amber_forceField} and CHARMM~\cite{MacKerell1998_charmm_forceField}, without explicit quantum-mechanical~(QM) calculations. This approach affords high computational throughput but inherently cannot faithfully capture quantum phenomena, including charge transfer, bond making/breaking, and many-body polarization. On the other end, \emph{ab initio} MD~(AIMD) treats the electronic structure with QM methods, yielding quantum-level accuracy, but with substantially higher computational cost, which limits its applicability. In recent years, machine-learning interatomic potentials (MLIPs) that approximate the QM potential energy surface~\cite{Fiedler2022_ml_potentials_review, Jacobs2025_machine_learning_potentials}, often implemented as neural network potentials (NNPs), have shown to be capable of bridging the accuracy–efficiency gap. These models typically employ deep or graph neural networks trained on density functional theory (DFT) reference data to yield near-\emph{ab initio} energies and forces via fast inference, substantially reducing per-step cost relative to direct DFT. As demand for quantum-level fidelity at scale grows, integrating machine-learning potentials in GROMACS becomes a natural step to extend the package features and provide a high-performance, accurate MD workflow.

Several software frameworks have emerged to construct and deploy MLIPs, facilitating their integration into established classical MD engines. Nevertheless, MLIPs are often not natively compatible with software engineering solutions commonly employed in classical MD, particularly domain decomposition on distributed-memory architectures. As a consequence, ML-driven simulations are frequently restricted to single process executions, imposing stringent constraints on the available computational power and memory capacity. The integration of MLIP models into mature MD engines without intrusive code refactoring is non-trivial and often relies on ad hoc solutions.

In this work, we extend the existing interface for MLIPs in GROMACS, called \texttt{NNPot}, by coupling the DeePMD-kit framework~\cite{zeng2025deepmdkitV3} with full support for domain-decomposed, GPU-accelerated simulations. The implementation builds on the existing GROMACS infrastructure to enable fast MD with near-\emph{ab initio} accuracy across multi-GPU and multi-node systems, while preserving the standard, feature-complete workflow familiar to users. To validate the coupling, we construct and in-house train a DeePMD model \mbox{(DPA-1)} and benchmark performance on systems equipped with NVIDIA A100 and AMD MI250x GPUs. We further analyze execution behavior using the \texttt{ROCm System Profiler} to identify principal performance bottlenecks and future optimization opportunities.

The contributions of this work include:
\begin{itemize}
    \item We integrate the MLP framework DeePMD-kit into the classical MD engine GROMACS, using a domain decomposition layer decoupled from the one employed by the main simulation, enabling scaling across multiple GPUs on different nodes.
    \item We build and in-house train a DPA-1 model of the DeePMD family that is used to evaluate the performance of the GROMACS-DeePMD integration.
    \item We conduct a comprehensive performance analysis of the code on NVIDIA A100 and AMD MI250x GPUs via scaling tests, and we perform a detailed code profiling using the \texttt{ROCm System Profiler} to highlight the implementation bottlenecks.
\end{itemize}

The remaining sections are organized as follows. In Sec.~\ref{sec_background}, we describe the MD method, MLIPs, DeePMD-kit, and the GROMACS code. In Sec.~\ref{sec_related_work}, we review related work. We present the methods and implementation in Sec.~\ref{sec_methodology}. The hardware and software setup employed in the tests is detailed in Sec.~\ref{sec_experimental_setup}, while we present the results in Sec.~\ref{sec_results}. Sec.~\ref{sec_discussion_conclusion} is dedicated to discussion and conclusion.

\section{Background}
\label{sec_background}
\subsection{Classical and ab initio molecular dynamics}

In \textbf{classical MD simulations}, atomic interactions are modeled within a purely classical framework, without explicit QM calculations~\cite{Allen2017computerSimulationOfLiquids, ZhouLiu2022moleculaDynamicsBook}. The interactions are described by a classical potential energy function, $E(\mathbf{r})$, whose functional form and parameters are provided by empirical force fields such as AMBER~\cite{Cornell1995_amber_forceField} and CHARMM~\cite{MacKerell1998_charmm_forceField}, obtained by fitting the analytic expression of $E(\mathbf{r})$ to QM results and experimental data.

The potential energy function is typically decomposed into terms that describe different classes of interactions independently, as shown in Eq.~\ref{eq_energyTot}:
\begin{equation}
    E_{\text{total}} = E_{\text{bonded}} + E_{\text{non-bonded}}^\text{short-range} + E_{\text{non-bonded}}^\text{long-range}.
    \label{eq_energyTot}
\end{equation}
Bonded interactions include contributions from bonds and angles (usually modeled as harmonic oscillators) and dihedrals, which depend on torsion about a bond, while out-of-plane deformations are represented via improper dihedrals. Non-bonded short-range interactions comprise Van der Waals forces, typically modeled by the Lennard-Jones (LJ) potential. Long-range electrostatic interactions are described by a Coulomb potential and are usually evaluated using the particle-mesh Ewald (PME) method. The Coulomb term is divided into a fast decaying real-space contribution and a reciprocal-space contribution. The former is computed through pairwise interactions in real space and, together with the LJ term, contributes to $E_{\text{non-bonded}}^{\text{short-range}}$. The latter is obtained by spreading charges onto a Cartesian mesh and solving the Poisson equation in Fourier space via three-dimensional FFTs, yielding $E_{\text{non-bonded}}^{\text{long-range}}$.

Once the potential is defined, the forces acting on the \mbox{atoms $i$} are computed by differentiating the energy with respect to the atomic positions $\mathbf{r}_i$, as shown in Eq.~\ref{eq_forceGradE}:
\begin{equation}
    \mathbf F_i = -\nabla_{\mathbf r_i} E.
    \label{eq_forceGradE}
\end{equation}
Atomic trajectories are then propagated in phase space by numerically integrating Newton's equations of motion, typically using leap-frog or Verlet algorithms.

It is instructive to outline how energies and forces are evaluated at the atomic level. If the total potential energy $E$ can be written as a sum of pairwise contributions $\phi_{ij}$, where $i$ and $j$ index atoms, then Eq.~\ref{eq_totE_pairwise} holds:
\begin{equation}
    E = \sum_{i}\sum_{j<i} \phi_{ij},
    \label{eq_totE_pairwise}
\end{equation}
where the restriction $j<i$ corresponds to the half-list convention, avoiding double counting pair interactions.
Then, the force on atom $i$ follows by differentiation, as in Eq.~\ref{eq_force_pairwiseE}:
\begin{equation}
    \mathbf F_i = -\sum_{j\neq i} \frac{\partial \phi_{ij}}{\partial \mathbf r_{i}}
    = -\sum_{j\neq i} \frac{\partial \phi_{ij}}{\partial r_{ij}}\frac{\partial r_{ij}}{\partial \mathbf r_i}, \;\; r_{ij} = ||\mathbf r_i - \mathbf r_j||.
    \label{eq_force_pairwiseE}
\end{equation}
Alternatively, the total energy $E$ can be expressed as a sum of per-atom (virtual) energies $e_i$, obtained by splitting each pair contribution equally between the two atoms involved, as in Eq.~\ref{eq_totE_singleAtomE}:
\begin{equation}
    E = \sum_i e_i, \;\; e_i = \frac{1}{2} \sum_{j\neq i} \phi_{ij}, 
    \label{eq_totE_singleAtomE}
\end{equation}
In this formulation, forces are computed by differentiating with respect to the atomic coordinates, as reported in Eq.~\ref{eq_forceAtomicEnergy}:
\begin{equation}
    \mathbf F_i = -\nabla_{\mathbf r_i} E 
    = -\sum_{j} \frac{\partial e_j }{\partial \mathbf r_{i}}
    = -\frac{\partial e_j }{\partial \mathbf r_{i}} -\sum_{j\neq i} \frac{\partial e_i}{\partial \mathbf r_{i}}.
    \label{eq_forceAtomicEnergy}
\end{equation}

The cost of computing all pairwise interactions among $N$ particles scales as $\mathcal O(N^2)$. To reduce this cost, MD simulations evaluate non-bonded short-range interactions only for atom pairs within a cutoff distance of a given atom $i$, i.e., within the neighbor list $\mathcal{N}(i)$. This reduces the computational complexity from quadratic to linear in $N$. The cost of performing an FFT on a Cartesian grid with $N_g$ grid points scales as $\mathcal{O}(N_g\log N_g)$, therefore, the overall computational cost for a classical MD simulation scales as $\mathcal{O}(N)+\mathcal{O}(N_g\log N_g)$. \\

\textbf{Quantum-accurate modeling} of chemical structures can be achieved with \textit{ab initio} molecular dynamics, where forces are computed on the fly from a QM treatment of the electronic structure~\cite{Iftimie2005_abInitio_md}. AIMD employs either the Born–Oppenheimer theory or the Car–Parrinello method and provides higher accuracy and fidelity than classical MD, as it naturally captures bond-breaking/forming, electronic polarization, and charge transfer. This accuracy comes at a higher computational cost, typically scaling as $\mathcal{O}(N^3)$ with system size, although linear-scaling DFT methods exist and can be practical for specific classes of systems~\cite{schade2022_aimd_linear}. In standard biomolecular simulations, systems are typically limited to a few thousand atoms and relatively short trajectory lengths.

\subsection {DeePMD-kit \& Deep Potentials architectures}
\label{sec_deepmd}

In recent years, MLIPs have become increasingly popular~\cite{noe2020_mlp_review} because they address the accuracy-cost trade-off between classical MD and AIMD. They approximate the Born-Oppenheimer potential energy $E(\mathbf{r})$ by learning from large QM datasets of energies and forces. Unlike empirical force fields with predefined functional forms, MLIPs employ flexible function approximators that better capture complex interactions and quantum effects. Several atomic encodings and model families have been developed, including kernel-based methods, basis-expansion approaches, deep neural networks, and graph neural networks. Once trained, MLIPs can reach near-QM accuracy and, with computational cost scaling as $\mathcal{O}(N)$, enable large and long MD simulations with near \textit{ab initio} fidelity.\\

\textbf{DeePMD-kit} (Deep Potential Molecular Dynamics kit)~\cite{wang2018deepmdkitV1} is an open-source framework for integrating Deep Potential family MLIPs (\emph{Deep Potentials}, DP) into MD engines. It offers commands and APIs covering the full DP workflow, including dataset construction, training, compression, and deployment for production MD.
Version 3~\cite{zeng2025deepmdkitV3} introduces a multi-backend architecture supporting TensorFlow, PyTorch, JAX, and PaddlePaddle for training and inference, with model backend conversion where available to enable flexible deployment without costly retraining. The framework has unified Python and C/C++ APIs for training and inference. 

DeePMD-kit accelerates training and inference via shared- and distributed-memory parallelism: multi-process training is natively supported for TensorFlow, PyTorch, and Paddle, while distributed-memory inference relies on domain decomposition provided by the host MD engine. Thread-level parallelism is controlled through environment variables. GPU acceleration is supported for both training and inference, builds are configured at compile time and can target NVIDIA CUDA or AMD ROCm/HIP.\\

\textbf{Deep potential models} do not substitute the entire computation of energies and forces in MD simulations. They encode only the information of each atom and its local environment (i.e., atoms within the neighbor list) into a descriptor, which is then used to compute atomic energies and short-range forces. Since the models are trained end-to-end to map atomic configurations to total energies and forces, they do not decompose the potential into explicit bonded, angular, dihedral, or non-bonded components. Rather, they provide a single unified energy/force term that implicitly captures all the interactions within the cutoff, 
thus yielding a more natural, data-driven representation of interatomic interactions.

\begin{figure}[!h]
    \centering
    \includegraphics[width=0.75\linewidth]{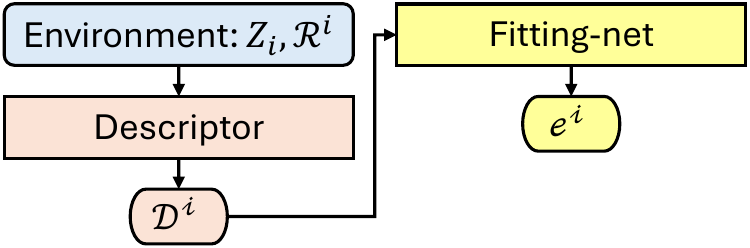}
    \caption{General deep model architecture. $Z_i$ denotes the atom type, $\mathcal{R}^i$ atom positions, $\mathcal{D}^i$ the atom descriptor, and $e_i$ the atom energy.}
    \label{fig_dp_architecture}
\end{figure}

The DeePMD team has developed an entire family of DP models that follow the so-called \textit{descriptor} plus \textit{fitting-net} architecture, as depicted in Fig.~\ref{fig_dp_architecture}. The descriptor is a symmetry preserving operator that embeds the information (position and type) of atom \textit{i} (called \textit{center}) and its neighboring atoms (\textit{edges}) into a local representation $\mathcal{D}^i$. It enforces translational, rotational, and (within each species) permutational symmetries. The fitting-net is a conventional multi layer perceptron (MLP) that maps $\mathcal{D}^i$ into the corresponding atomic energy $e_i$, $e_i=\Phi(\mathcal{D}^i)$. The total system energy is then $E = \sum_i e_i$, and since forces are conservative by construction, they can be computed using automatic differentiation as energy gradients, with Eq.~\ref{eq_forceGradE}. For this work, the relevant architectural distinction is between local
single-cutoff descriptors, evaluated within one neighbor list, and message-passing descriptors, whose dependencies expand over multiple hops.

\begin{figure*}[!h]
    \centering
    \subfloat[\label{fig_dpSe}]{\includegraphics[width=0.23\linewidth]{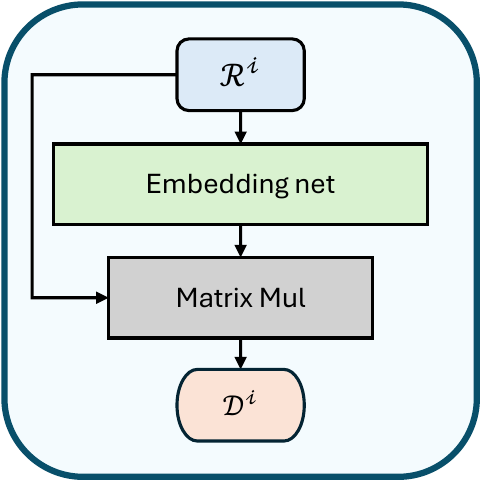}}
    \hspace{1pt}
    \subfloat[\label{fig_dpa1}]{\includegraphics[width=0.23\linewidth]{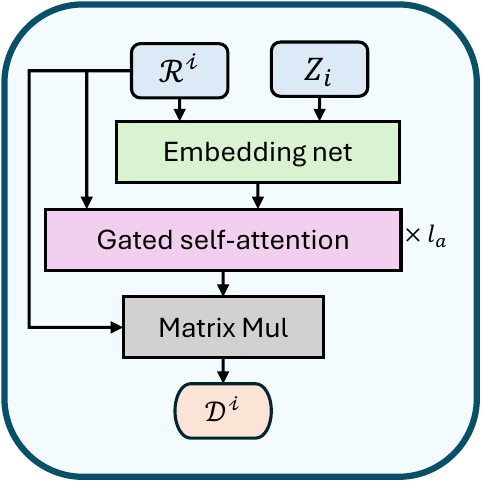}}
    \hspace{1pt}
    \subfloat[\label{fig_dpa2}]{\includegraphics[width=0.23\linewidth]{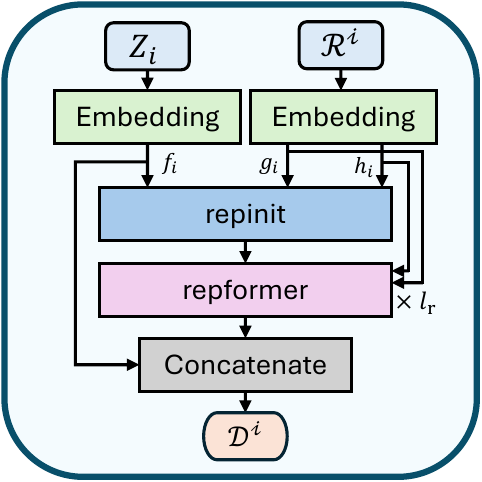}}
    \hspace{1pt}
    \subfloat[\label{fig_dpa3}]{\includegraphics[width=0.23\linewidth]{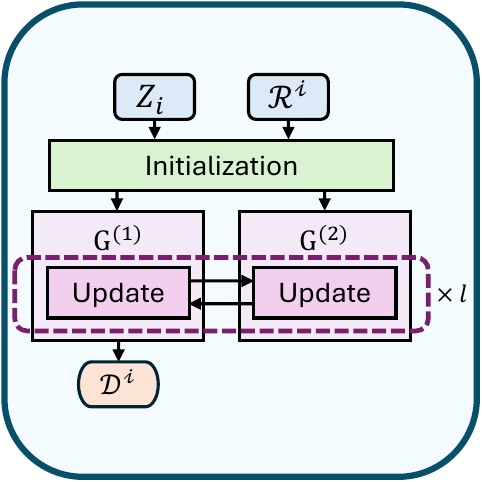}}
    \caption{DP-SE (a), DPA-1 (b), DPA-2 (c), and DPA-3 (d) descriptor architectures. $Z_i$ denotes the atom type, $\mathcal{R}^i$ atom positions, and $\mathcal{D}^i$ the atom descriptor. The standard DPA-3 configuration (d) uses a graph series of order 2.}
\end{figure*}

There exist four major DP model classes, characterized by different descriptor architectures. Deep Potential - Smooth Edition (\textbf{DP-SE})~\cite{zhang2018DeepPotSE} is the first DP model developed. The descriptor is built by combining a local environment matrix $R^i$, describing neighbor geometry in invariant coordinates, with a feature matrix $G^i$ (and reduced-dimensional $G^i_r$) generated by a multi-layer embedding network: $\mathcal{D}^i=(G^i)^T R^i (R^i)^TG^i_r$. The architecture is reported in Fig.~\ref{fig_dpSe}. 

\textbf{DPA-1}~\cite{zhang2022dpa1}, described in Fig.~\ref{fig_dpa1}, enhances the DP-SE descriptor by introducing a gated self-attention operator and atom type embeddings $Z$, enabling pretraining across multiple chemistries and subsequent finetuning for specific tasks.
Starting from the DP-SE pipeline, DPA-1 includes the type embedding into the feature matrix $G^i$ and adds $l_a$ gated self-attention blocks that operate within the neighbor set of the center $i$. The gate mask introduces information regarding the angular correlation $R^i(R^i)^T$. The self-attention operation combines the information of the atoms in the neighbor list, and has been shown to improve accuracy relative to DP-SE across diverse systems.
As in DP-SE, descriptors are computed per center atom using only its local neighborhood. Edges are fixed during the evaluation, the attention operates only over neighbors of the same center $i$ and does not introduce inter-center coupling. Thus, the architecture maintains the locality of the model, which is a key aspect in distributed-memory MD simulations.

\textbf{DPA-2}~\cite{zhang2023dpa2}, depicted in Fig.~\ref{fig_dpa2}, generalizes the descriptor using three coupled channels to represent each atom and its local environment, encoding single-atom features, pair invariants, and equivariants. A representation-initializer layer (\texttt{repinit}) generates the first embedding, then, a stack of \texttt{repformer} layers updates the representation, each layer comprising multi-head self-attention and multi-head gated attention. In contrast to DPA-1, DPA-2 introduces message passing on the neighbor graph: after each \texttt{repformer} layer, the updated center representation of atom $i$ is propagated to all connected neighbors, updating the corresponding edge features. This enables information to propagate beyond the predefined neighbor list, capturing longer-range interactions and higher-order many-body correlations. With $l_r$ \texttt{repformer} layers, a center representation depends on atoms within a distance $r_c \times l_r$.

\textbf{DPA-3}~\cite{zhang2025dpa3}, shown in Fig.~\ref{fig_dpa3}, is based on a graph neural network (GNN). Its descriptor is a multi-layer message-passing GNN defined on a Line-Graph Series (LiGS) of order $K$, with $G^1$ the initial graph. As in DPA-2, the key point for MD-engine integration is that each layer propagates information by one additional hop on the neighbor graph, thus, after $l$ layers, a node’s representation depends on all atoms within $l$ hops on $G^1$.

DPA-2 and DPA-3 are termed \textit{Large Atomic Models} (LAMs) to emphasize their ability to represent atomic configurations across diverse tasks.
In this paper, as detailed in Sec.~\ref{sec_methodology}, we focus on local models such as DP-SE and DPA-1. \mbox{DPA-2} and \mbox{DPA-3} message-passing architectures are discussed to provide a complete overview of model families and to make explicit the different MD-engine integration requirements they introduce.

\subsection{Neighbor list \& Domain decomposition simulations}
\label{sec_DD_simulation}
When integrating DP models into classical MD engines several aspects must be considered, particularly regarding neighbor lists calculation and strategies for distributed memory simulations.\\

In the most straightforward approach, \textbf{neighbor lists} are constructed as full lists: for each atom $i$, the set $\mathcal{N}(i)$ contains all atoms within the cutoff distance of $i$. Since forces in classical MD decompose into pairwise interactions and satisfy $F_{AB} = -F_{BA}$ (Newton’s third law), storing full-list entries for both directions is redundant. A more efficient alternative is to use so-called half lists, in which $\mathcal{N}(i)$ includes only atoms $j \mid j<i$. In this way, each pair interaction is counted exactly once. While this strategy is effective in classical MD software~\cite{pall2013gromacs_verlet}, it is not suitable for DP models, since constructing the atomic descriptor $\mathcal{D}^{i}$ requires information about the entire local environment of atom $i$. Therefore, full neighbor lists are needed when DP models are employed.\\

In \textbf{domain-decomposition (DD) simulations}, the spatial domain is partitioned among ranks, and each rank integrates the equations of motion for the local atoms. To evaluate pair forces between atoms located on different subdomains, information must be exchanged across ranks. A widely used strategy is the \textit{half-shell} method: each rank extends its subdomain with a \textit{halo} (ghost layer) of thickness equal to the cutoff $r_c$ that contains ghost copies of atoms owned by neighboring ranks. Pair interactions between a local atom $i$ and its ghost neighbors $j$ are computed only once on one of the two ranks, usually the one that owns $i$ if a half-list is used. The resulting force contributions $\mathbf{F}_{ij} = -\mathbf{F}_{ji}$ are accumulated locally and then communicated to the ranks owning atoms $j$ so that the forces on the real atoms $j$ are updated.

Two additional strategies further reduce the amount of inter-rank communication. In the \textit{eighth-shell} method~\cite{Liem1991_DD_eighthShell}, a pair interaction $(i,j)$ may not be computed on the home rank of $i$ or $j$. Instead, it is evaluated on a third \textit{interaction rank}, chosen by minimizing the Cartesian indices of the owning ranks, selecting one of the eight octants of the domain.
Similarly, the \textit{midpoint} method~\cite{Bowers2006_dd_midpoint} assigns the interaction $(i,j)$ to the rank whose box contains the midpoint of the two atoms, i.e., the point $(\mathbf{r}_i+\mathbf{r}_j)/2$. In both methods, the pair forces $\mathbf{F}_{ij}$ are accumulated on the interaction rank and then communicated to the ranks owning atoms $i$ and $j$.\\

\begin{figure}[!h]
    \centering
    \includegraphics[width=0.8\linewidth]{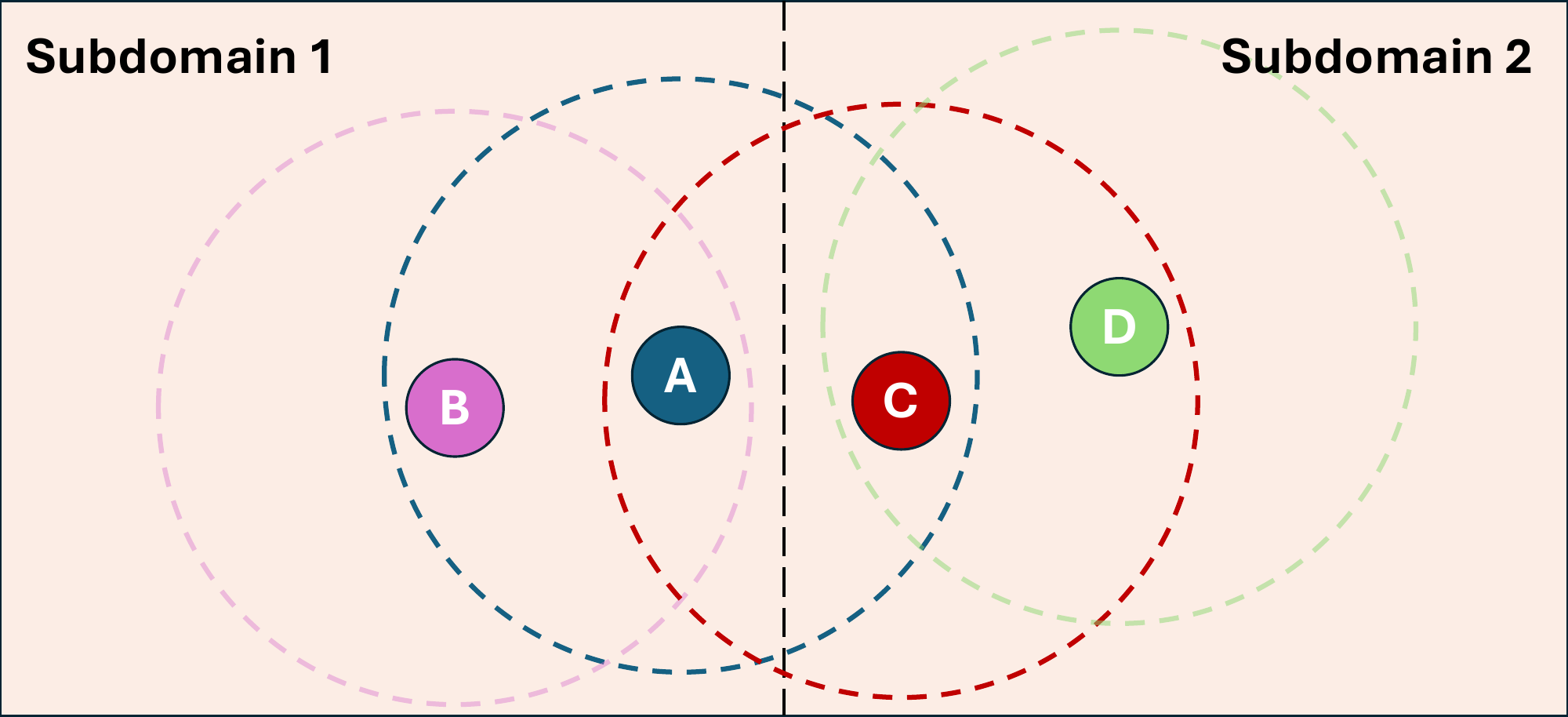}
    \caption{Neighbor list in case of domain decomposition. Atoms 'A' and 'B' are local in subdomain 1, while atoms 'C' and 'D' belong to subdomain 2.}
    \label{fig_neighbor_list_dd}
\end{figure}

Using Deep Potential models in domain-decomposition (DD) simulations requires special care. First, constructing the atomic descriptor $\mathcal{D}^i$, and hence computing the atomic energy, demands the complete full neighbor list $\mathcal{N}(i)$. Since the force on atom $i$ is obtained by differentiating the atomic energies of all atoms in $\mathcal{N}(i)$ (Eq.~\ref{eq_forceAtomicEnergy}), all atomic energies must be evaluated consistently, including those of neighboring atoms. Thus, for every neighbor $j \in \mathcal{N}(i)$, one must also assemble the full list $\mathcal{N}(j)$ to compute the correct descriptor $\mathcal{D}^j$. In a DD simulation with a basic half-shell communication strategy, this would require constructing a halo of thickness $2r_c$ to import the additional ghost atoms needed to complete the neighbor lists of the first layer of ghost atoms ($\{\mathcal{N}(j)\mid j\in\mathcal{N}(i)\}$). Fig.~\ref{fig_neighbor_list_dd} illustrates the atom dependencies in the case of DD. The neighbor list of atom 'A', local in subdomain 1, includes atoms 'B' and 'C'. Since atom 'D' $\in \mathcal{N}(C)$, computing the correct force on 'A' would require the information of 'D' as well.

The DeePMD-kit \texttt{compute} API  allows the caller to flag which atoms are local and which are ghosts. Then, during force evaluation, contributions from ghost neighbors are neglected, as shown in Eq.~\ref{eq_forcesLocalGhost}:
\begin{equation}
\mathbf F_i = -\sum_{j\in \mathcal{N}(i)} m_j \frac{\partial E_j}{\partial \mathbf r_i}, \; \; m_j = \begin{cases}
1,& j \; \; \text{ is local} \\
0,& j\; \; \text{ is ghost}
\end{cases}.
\label{eq_forcesLocalGhost}
\end{equation}
Forces on ghost atoms $j$ (-$\partial E_i/\partial \mathbf r_j$) are still evaluated and then communicated to the owning ranks, where they are added to the corresponding real-atom force accumulators. Thus, the standard symmetric communication scheme used in the half-shell approach with an $r_c$ halo can be used.

This approach works well for DP-SE and DPA-1 models, where all information used by the atomic descriptor is confined to the local neighbor list within a single cutoff. However, it breaks for message-passing models such as \mbox{DPA-2} and \mbox{DPA-3}, where, as discussed in Sec.~\ref {sec_deepmd}, with $l$ message-passing layers, the descriptor $\mathcal{D}^i$ for atom $i$ depends on atoms as far as $l$ cutoffs away. A straightforward way to enable DD simulations is therefore to enlarge the half-shell halo to a thickness of $r_c\times (l+1)$, but this approach significantly increases the computational cost. 
To handle message-passing models without constructing extended ghost-of-ghost neighbor lists, DeePMD-kit v3 has introduced a custom C++ communication operator that uses MPI to exchange atom features across ranks during inference. Additionally, DeePMD-kit has a lightweight C++ struct (\texttt{InputNlist}) that wraps the information required by the custom communicator. \texttt{InputNlist} takes as input the full neighbor lists of the local atoms, which must already include the first layer of ghost atoms, along with information regarding ownership of ghost atoms by other ranks. Additionally, \texttt{InputNlist} consumes an MPI communicator provided by the host MD engine. During the evaluation of a multi-layer message-passing model, DeePMD alternates computation and MPI exchanges, ensuring that ghost atom features remain synchronized as information propagates layer by layer.

In DD simulations, all DP models require each rank to have at least one halo layer of ghost atoms to construct full neighbor lists for the local atoms. When DD is implemented with the eighth-shell or midpoint approaches, this guarantee does not hold, since some ranks may not create ghost atoms. Therefore, the communication patterns of the eighth-shell and midpoint schemes are not suitable for a straightforward deployment of DP models in distributed memory simulations.

\subsection{GROMACS}
\label{sec_gromacs}
GROMACS~\cite{pronk2013_gromacs,abraham2015gromacs,pall2020gromacs,gromacsManual} is a C/C++ open-source classical MD engine, designed for performance portability across a wide range of hardware and system sizes. It employs multi-level parallelism, where shared-memory parallelism is implemented with OpenMP threading and distributed-memory parallelism with the MPI standard. CPU calculations are further optimized by leveraging low-level SIMD instructions. GPU acceleration is available through CUDA, HIP, and SYCL backends. When using GPUs, GROMACS supports GPU-aware MPI and GPU-direct communication, as well as vendor GPU communication stacks such as NVSHMEM~\cite{doijade2025_gromacs_nvshmem}, NCCL, and RCCL. 

\begin{figure}[h!]
    \centering
    \includegraphics[width=0.99\linewidth]{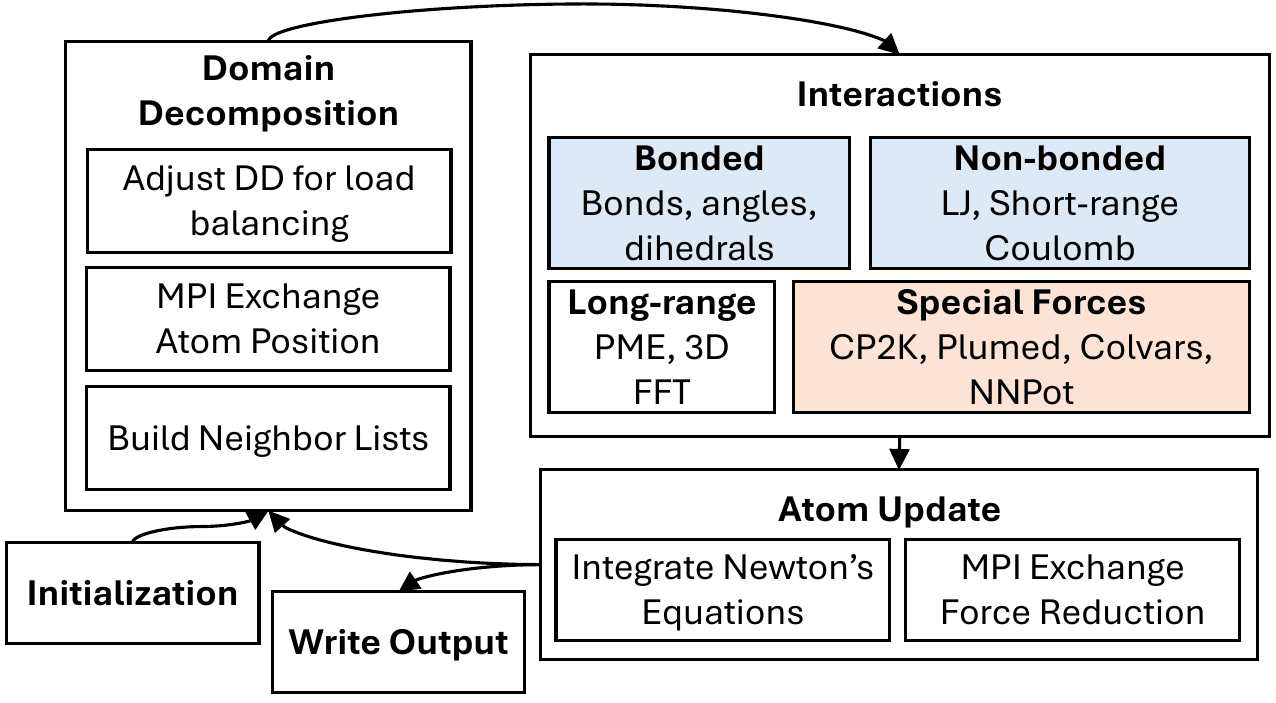}
    \caption{GROMACS MD simulation main loop. The conceptual step order is: (1) initialization, (2) domain decomposition and load balancing, (3) position exchange, (4) neighbor-list construction, (5) interaction evaluation, (6) special-force, (7) force reduction and atom update, and (8) output. The actual implementation pipelines several of these stages.}
    \label{fig_gromacs_md_main_loop}
\end{figure}

In standard MD runs, GROMACS employs empirical force fields to compute bonded and short-range atom energies and forces, while long-range interactions (Coulomb) are calculated with the PME approach. Additionally, GROMACS provides a specific module, \textit{special forces}, that enables the inclusion of extra interactions or external couplings within a simulation. Currently, this module has dedicated interfaces for CP2K (for QM/MM runs), Plumed, and Colvars. The GROMACS 2025 release has introduced an additional interface, named \texttt{NNPot}, which is used to include custom machine learning potentials with a PyTorch backend. Fig.~\ref{fig_gromacs_md_main_loop} details the structure of the main MD simulation loop in the software. In the actual build, the computational steps are reorganized and pipelined to maximize overlap between computation and communication, as well as between CPU and GPU work. Thus, the execution order depends on the hardware configuration and may differ from the schematic shown in Fig.~\ref{fig_gromacs_md_main_loop}.

GROMACS uses half neighbor lists built by a highly optimized algorithm \cite{pall2013gromacs_verlet}, and it implements an advanced eighth-shell scheme to minimize communication in DD simulations. These algorithmic choices deliver high performance for classical MD simulations, however, they complicate the integration of DP models, as explained in Sec.~\ref{sec_DD_simulation}. In the current implementation, when the \texttt{NNPot} module is used under DD, all atomic data are exchanged across ranks via an MPI all-to-all communication, after which only \mbox{process 0} performs the inference and then distributes the resulting forces to the other ranks via another MPI collective call. This design effectively collapses the DD run into a single-domain workflow, eliminating the issues related to domain halos and ghost atoms. However, this choice currently limits the size of systems that can be simulated, as the inference task can leverage only the computational power and memory of a single process. This limitation is removed in the present work via distributed-memory inference across ranks.

\section{Related Work}
\label{sec_related_work}
In recent years, several MLIPs frameworks, such as SNAP~\cite{thompson2015_snap}, SIMPLE-NN~\cite{Lee2019_simpleNN}, ANET~\cite{artrith2016aenet}, ANI~\cite{Gao2020torchANI}, TorchMD~\cite{doerr2021torchmd,pelaez2024torchmd-net2}, NequIP~\cite{Batzner2022_NequIP}, Allegro~\cite{Musaelian2023allegro_lammps100M}, MALA~\cite{cangi2025_mala}, and the aforementioned DeePMD-kit~\cite{wang2018deepmdkitV1} have been developed. Now, many MD codes integrate MLIPs via native modules or by leveraging third-party packages. For instance, the CHARMM suite~\cite{Hwang2024charmm} has introduced the support for ML potentials via a built-in extension in the new PyCHARMM toolkit, while DeePMD-kit has been integrated into the Amber software~\cite{Zeng2021_amber_deepmdkit, Case2023ambertools} and the OpenMM engine~\cite{ding2024_openmm_deepmdkit}. The MD code LAMMPS~\cite{thompson2022lammps} has been coupled with a broad ecosystem of ML methods: it has official support for the \mbox{DeePMD-kit}~\cite{zeng2025deepmdkitV3}, while other ML models can be incorporated with packages like AENET-LAMMPS~\cite{chen2021aenetLAMMPS}, MLMOD~\cite{atzberger2023_mlmod_lammps}, and chemtrain-deploy~\cite{fuchs2025chemtrain}. LAMMPS implements a half-shell scheme with symmetric ghost halos in the case of DD simulations, and it has options to enable full neighbor lists. Thus, it is inherently suitable to integrate DP models and run on memory-distributed systems. DD simulations have been run coupling LAMMPS with the Allegro~\cite{Musaelian2023allegro_lammps100M} and the SNAP~\cite{NguyenCong2021_snap_lammps1B} packages. 
Additionally, LAMMPS coupled with DeePMD-kit has been used to perform large simulations on Fugaku and Summit supercomputers~\cite{weile2020deepMD_lammps_gordonbell_100milion,Guo2022_deepmd_lammps_3B,Li2024_deepmd_lammps_149ns}.
The custom C++ communicator recently added to DeePMD-kit V3~\cite{zeng2025deepmdkitV3} for message-passing DP models, described in Sec.~\ref{sec_DD_simulation}, has been developed specifically for integration with LAMMPS.
DeePMD-kit provides an official patch for integration with GROMACS. The patch targets GROMACS 2020.2 and is not compatible with more recent GROMACS versions. Additionally, it restricts the simulations to a single domain.

\section{Methodology}
\label{sec_methodology}
\subsection{DeePMD-kit Integration}
\label{sec_methodology_deepmd_int}
As detailed in Secs.~\ref{sec_DD_simulation} and \ref{sec_gromacs}, the highly optimized GROMACS implementation is not inherently suitable to integrate DP models in a way that supports distributed memory simulations. Achieving a robust coupling with DeePMD-kit (akin to LAMMPS) would require a substantial modification of GROMACS's domain decomposition architecture, including the introduction of per-subdomain halos and symmetric ghost-atom exchanges. Such changes would imply invasive code refactoring that lies well beyond the scope of this work and might even hinder the current performance of the software.

In this work, we leverage the existing \texttt{NNPot} interface, extending it to integrate DeePMD-kit and to enable DD simulations. \texttt{NNPot} already provides the routines required to preprocess the atomic structures for which forces are evaluated by the DP model. Specifically, during the preprocessing stage, executed prior to the MD run, these routines modify the molecular topology by removing all bonded interactions and by adding the marked atoms to the \emph{exclusion lists}. The marked atoms are omitted from the neighbor lists and no short-range interactions are computed for them. Long-range (Coulomb) interactions are evaluated as usual. During the MD run, at each time step, only the marked atoms, hereafter called NN atoms, are passed to \texttt{NNPot}, which performs the model inference.

\begin{figure}[h!]
    \centering
    \includegraphics[width=0.99\linewidth]{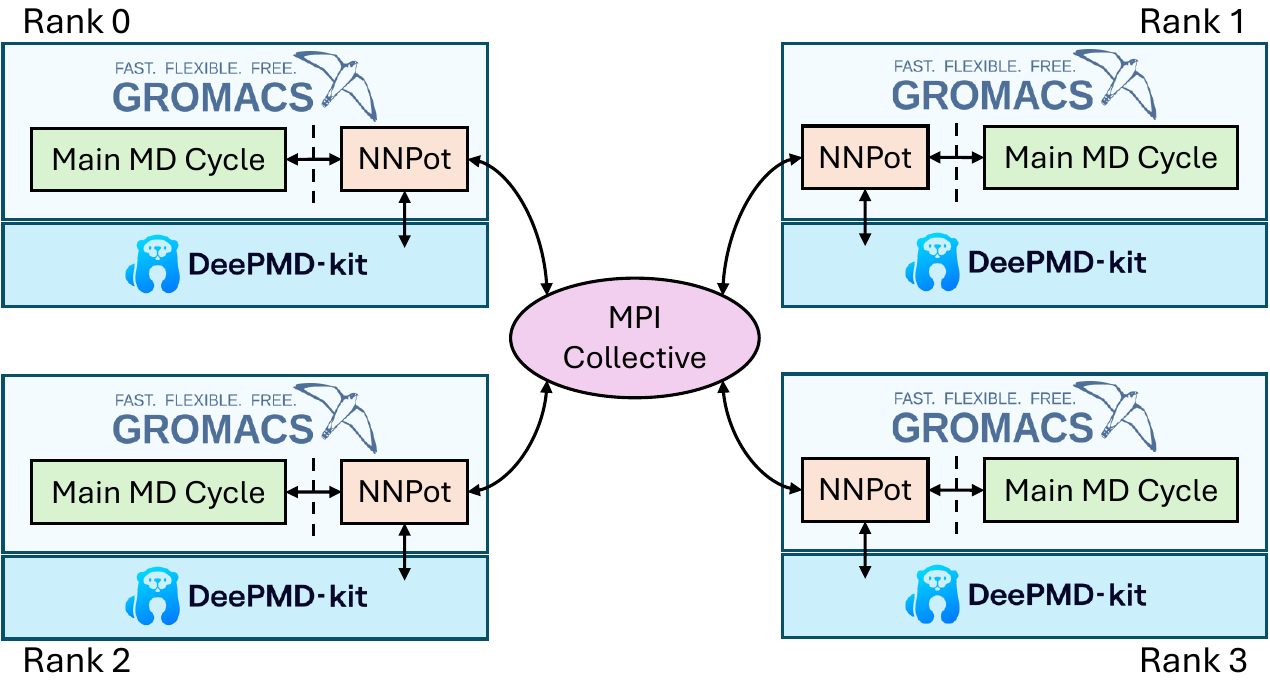}
    \caption{DeePMD-kit integration in the GROMACS MD engine in the case of DD simulations, example with four MPI ranks.}
    \label{fig_deepmd_gromacs_parallel}
\end{figure}

We integrate DeePMD-kit by introducing a dedicated \texttt{DeepmdModel} class within \texttt{NNPot}. This class holds a pointer to the Deep Potential (DP) model object, provides a wrapper around the \texttt{deepmd::compute()} API, and implements auxiliary routines for data-layout and unit conversions required before the inference.
Following the initial MPI collective call, every rank holds the positions of all NN atoms in the global system, stored in the vector \texttt{atomAll}. To enable distributed-memory simulations, a virtual domain decomposition is constructed: the simulation box is partitioned into a uniform Cartesian grid, and each rank extracts from \texttt{atomAll} only the atoms whose coordinates fall within its assigned subdomain (local atoms). In addition, each rank collects the atoms needed to build symmetric halo regions with ghost atoms, analogous to the half-shell scheme. Atom images required in the case of periodic boundary conditions are built as well. In this implementation, all ghost atoms required by the DP model to compute exact atomic descriptors are included, allowing the computation of correct forces on local atoms without the force-reduction stage.
Once the local and ghost atom buffers have been assembled, each rank forwards the corresponding subsystem to DeePMD-kit, which performs the inference. DeePMD-kit API is not MPI-aware, the inference is performed in parallel on independent ranks. After the inference stage, another MPI collective aggregates and redistributes the computed forces across all ranks. The integration of DeePMD in GROMACS is illustrated in Fig.~\ref{fig_deepmd_gromacs_parallel}.

We choose the DPA-1 architecture for this integration because its descriptor is strictly local depending only on atoms within a single cutoff neighborhood, without involving inter-center coupling (Sec.~\ref{sec_deepmd}). This locality allows each rank to independently compute exact descriptors and forces using only a $2r_c$ halo, as described in Sec.~\ref{sec_DD_simulation}, making DPA-1 naturally compatible with our virtual DD design. Message-passing models such as DPA-2 and DPA-3 would require enlarged halos of depth $(l+1)r_c$, in principle compatible with our virtual DD, but that would substantially increase the ghost-atom count and diminish the benefits of domain decomposition. This ''message-passing vs. halo growth'' trade-off is well recognized in the MLIP community and is one reason many scalable architectures keep message-passing depth small~\cite{Musaelian2023allegro_lammps100M,batatia2022mace}. Alternatively, DPA-2 and DPA-3 would require the DeePMD-kit's custom feature-exchange communicator based on full neighbor lists and a symmetric ghost layout akin to LAMMPS's DD infrastructure, currently not available in GROMACS. 

Our design trades memory for implementation simplicity. Replicating all NN atoms across all ranks might raise concerns regarding the memory footprint. We note that only the DP group (protein) is replicated, not the entire system, including the solvent. The per-rank memory footprint is linear in the number of NN atoms and consists of a small set of single-precision arrays (positions, types, indices), leading usually to a memory requirement of 28 bytes per NN atom. For typical biomolecular systems with 10,000 – 1,000,000 atoms, this corresponds to at most a few tens of MB per rank, and it is independent from the number of ranks. At very large device counts or for several-million-atom NN groups, a point-to-point halo exchange would be preferable.

The domain decomposition employed within \texttt{NNPot} is a temporary, \emph{virtual} decomposition that is entirely independent of the DD used by the main GROMACS MD loop. Atoms that are local to a given rank inside \texttt{NNPot} need not reside on the same rank during the main MD cycle. This virtual DD promotes a more uniform distribution of NN atoms across subdomains, which GROMACS does not generally guarantee due to its dynamic load balancing that takes into account all atoms in the system. The virtual DD is constructed by comparing atomic coordinates against the boundary coordinates of the subdomain assigned to each MPI rank, and halo regions are generated by expanding each subdomain by $2r_c$ in all Cartesian directions. Since no pairwise distances are evaluated, this procedure scales linearly with the number of atoms, $\mathcal{O}(N)$, and has a limited impact on overall performance.

\subsection{Deep Potential model architecture \& Training}
We employ a deep potential model based on the \mbox{DPA-1} architecture~\cite{zhang2022dpa1} to avoid message-passing limitations, as detailed in Sec.~\ref{sec_methodology_deepmd_int}. The model uses the \texttt{se\_attention\_v2} descriptor with three self-attention layers of hidden size 256. The embedding network has three layers with 32, 64, and 128 neurons, while the fitting network consists of three fully connected layers with 256 neurons each. The model has a total of 1.6 million parameters. We use PyTorch as the machine learning backend, and the model uses only single-precision (FP32) arithmetic.

The model is trained on a dataset of protein fragments solvated in water~\cite{Unke2019_physnet}. The dataset contains 2,594,609 unique frameworks, it is openly available at the \texttt{AIS Square} website\footnote{\url{https://www.aissquare.com/datasets/detail?pageType=datasets&name=Unke2019PhysNet_SolvatedProtein_DPA_v3_1}}, and has also been used by the DeePMD team to train the multi-task DPA-3 model.
We train the model for 2,000,000 epochs, which requires roughly 19 hours with a single NVIDIA GeForce RTX 4080 GPU card.

\section{Experimental setup}
\label{sec_experimental_setup}
\subsection{Hardware \& Software}
We test our implementation on two different systems. The first, here called System-1, is a Cray machine, in which each node is equipped with an AMD EPYC 7A53 64-core CPU, 512 GB of DDR4 RAM, and 4x AMD MI250x GPUs. Each AMD MI250x GPU features two graphical compute devices (GCDs), each with 64 GB of HBM2E memory, thus, each compute node can expose to the host eight devices.
The second machine, here referred to as System-2, features compute nodes equipped with 2x AMD EPYC Rome 7452 32-core CPUs and 512 GB of RAM, as well as 4x NVIDIA A100 GPUs. Each device has 40GB of HBM memory.
The software used in this work for the two systems is detailed in Tab.~\ref{tab_software}.

\begin{table}[h!]
    \centering
    \caption{Software experimental setup.}
    \label{tab_software}
    \begin{tabular}{c|cc}
    \toprule
    & System-1 & System-2 \\
    \midrule
    GROMACS   & v2025.2   & v2025.2   \\
    DeePMD-kit& v3.1.0    & v3.1.0    \\
    PyTorch   & v2.7.1    & v2.7.1    \\
    Compiler  & GCC 13.2.1 & GCC 13.3.0 \\
    SDK       & ROCm 6.3.3 & CUDA 12.6.0 \\
    MPI       & Cray-MPICH 8.1.31 & OpenMPI 5.0.3 \\
    \bottomrule
    \end{tabular}
\end{table}

\subsection{Protein simulations}
We select two different protein structures of different sizes to perform the tests. The first is the villin headpiece subdomain (PDB entry 1YRF\footnote{\url{https://www.rcsb.org/structure/1YRF}}), a protein constituted of 582 atoms. We employ this structure to assess the correctness of our in-house trained \mbox{DPA-1} model and the custom DD implementation by comparing DP results against a classical MD simulation. 
Then, we investigate the computational performance of the GROMACS-DeePMD coupling with the alpha actinin rod domain 1HCI\footnote{\url{https://www.rcsb.org/structure/1HCI}}, a larger protein structure. 1HCI is the central road domain of the human \mbox{$\alpha$-actinin} protein, and it is constituted by two antiparallel helicoidal atom chains. 1HCI has a total of 15,668 atoms.
For both protein systems, we employ the deep potential model only during the main MD run, whereas during the energy minimization (EM), NVT, and NPT equilibration stages, forces are computed with the standard CHARMM force fields. In all tests, the protein is solvated in a water solution with Na and Cl ions. Tab.~\ref{tab_proteinSimulations_param} details the simulation setups, reporting values for the time step ($\Delta$t), the number of steps, and the cutoff radius ($r_c$).   
\begin{table}[h!]
    \centering
    \caption{Protein simulations parameters.}
    \begin{tabular}{c|ccc|ccc}
    \toprule
         & \multicolumn{3}{c|}{Small Protein 1YRF} & \multicolumn{3}{c}{Large Protein 1HCI} \\
         \cmidrule(lr){2-4}\cmidrule(lr){5-7}
         & EM & NVT/NPT & MD & EM & NVT/NPT & MD \\
    \midrule
       $\Delta$t [fs]   & -  & 2     & 2   & -  & 2     & 2 \\
       Steps             & -  & 40,000 & 10,000 & -  & 40,000 & 200 \\
       $r_c$ [nm]      & 1.2& 1.2   & 0.8 & 1.2& 1.2   & 0.8 \\
       DP model           & No & No    & DPA-1 & No & No    & DPA-1 \\
       DP Group     & -  & -     & Protein & - & -  & Protein \\
    \bottomrule
    \end{tabular}
    \label{tab_proteinSimulations_param}
\end{table}

\subsection{GROMACS-DeePMD coupling validation}
We validate the implementation and our in-house trained DPA-1 model by simulating the 1YRF protein with two GPUs on two MPI processes and comparing the DeePMD run using the \mbox{DPA-1} model against a conventional force-field simulation. Both simulations are executed on System-2 equipped with NVIDIA GPUs. As a validation observable, we analyze the evolution in time of the protein radii of gyration about the Cartesian axes x, y, and z, which measure the structure anisotropy and its unfolding. A sustained and large increment in the gyration radii signals unphysical expansion of the protein, thus it would indicate an error in the model and/or DD implementation. GROMACS provides the radii of gyration with the \texttt{gyrate} command.

\subsection{Profiling setup \& Tools}
We evaluate the computational performance of the GROMACS–DeePMD coupling via strong and weak scaling tests on NVIDIA and AMD GPUs using the 1HCI protein, scaling up to 32 devices. We further profile an AMD-GPU run with 16 MPI processes using the \texttt{ROCm System Profiler}. This tool provides us a trace that captures HIP kernels, MPI calls, and, using the \texttt{roctx} APIs, DeePMD inference regions. In all simulations, we assign one GPU per MPI process and use eight OpenMP threads per rank.

For strong scaling, we simulate a single 1HCI chain while increasing the number of MPI ranks. For weak scaling, we increase the MPI processes and proportionally replicate the 1HCI system to maintain a constant load of eight MPI processes per protein (i.e., protein-to-processes ratio of 1:8). We report the code performance in scaling tests as nanoseconds per day (ns/day) using the GROMACS’s built-in timer. This is a standard MD metric for simulation speed that denotes the trajectory length achievable in 24 hours of wall-clock time (i.e., how many nanoseconds can be simulated per day).

\section{Results}
\label{sec_results}
\subsection{DPA-1 training \& Implementation validation}
\begin{figure}[!h]
    \centering
    \includegraphics[width=0.9\linewidth]{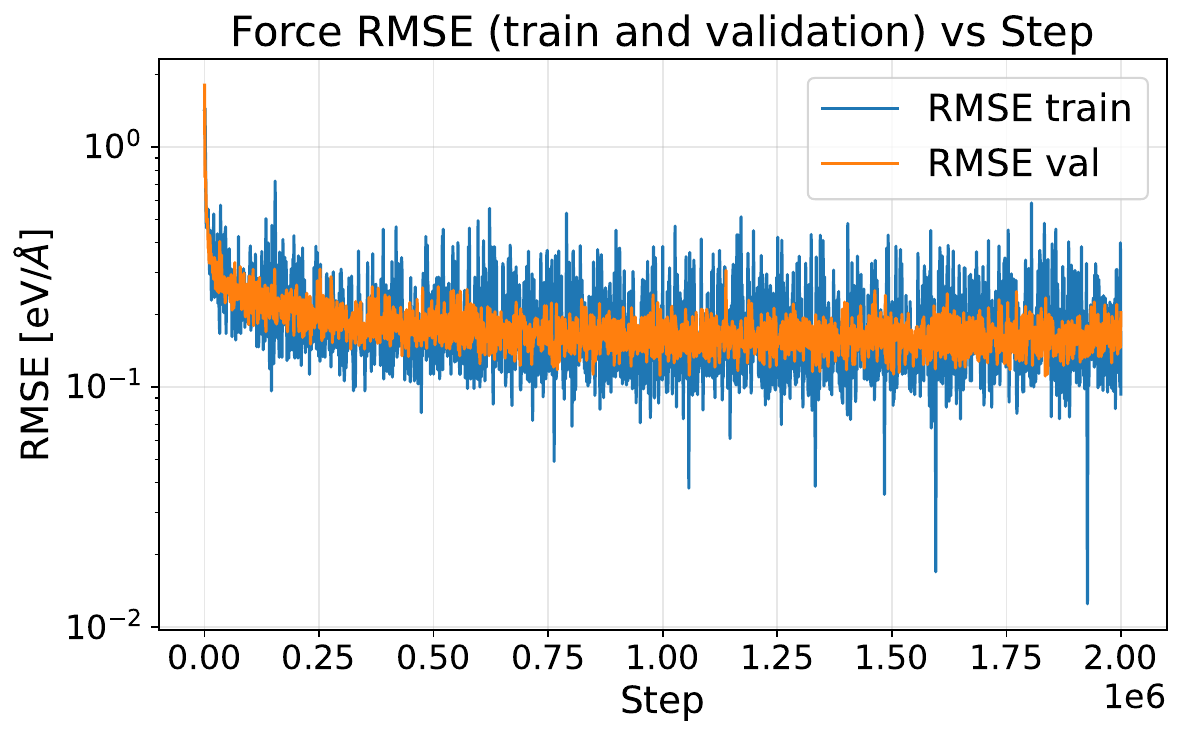}
    \caption{Evolution of the force RMSE during training of the DPA-1 model. The error on the training dataset is shown in blue, while the error computed against the validation dataset is reported in orange.}
    \label{fig_DPA1_train}
\end{figure}

We report in Fig.~\ref{fig_DPA1_train} the evolution of the force root mean square error (RMSE), calculated against the training and validation datasets, during the training of the DPA-1 model. The model reaches an error of roughly $0.2$ eV/\AA, which is aligned to the errors reported for the DPA-2 model~\cite{zhang2023dpa2}, after 750,000 steps, and then remains constant, a sign that we reached a plateau in the loss function. 

\begin{figure}[!h]
    \centering
    \includegraphics[width=0.9\linewidth]{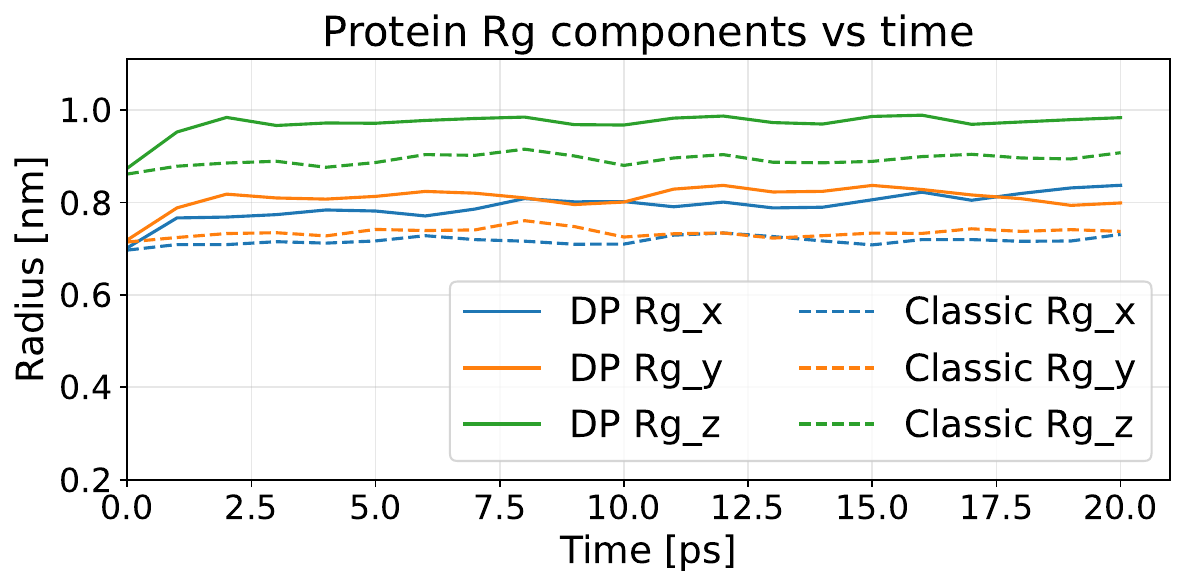}
    \caption{Comparison between the protein gyration radii about the three Cartesian axes x, y, z obtained with an MD simulation with the DPA-1 model (solid curves) and a classical MD simulation (dashed curves).}
    \label{fig_DPA1_benchmark}
\end{figure}

Fig.~\ref{fig_DPA1_benchmark} shows the evolution in time of the protein radii of gyration among the Cartesian axes x, y, z, comparing the \mbox{DPA-1} simulation with the standard force-field run. The three radii calculated using the DP model exhibit an offset of approximately 10\% with respect to the values obtained from the classical MD run. This is a reasonable behavior consistent with the \mbox{DPA-1} model having a different minimum in the potential energy surface. Most importantly, the radii remain stable over time, indicating a structurally stable protein (no "blowing up") and confirming the correctness of the domain decomposition in the GROMACS–DeePMD coupling.

\subsection{Performance analysis}
\begin{figure}
    \centering
    \includegraphics[width=0.99\linewidth]{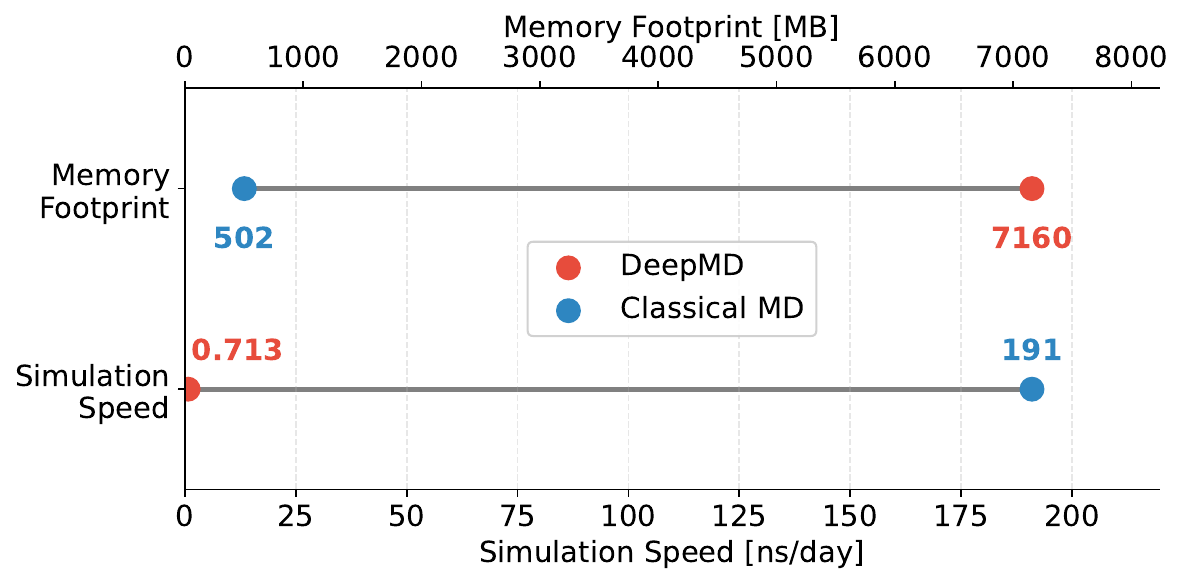}
    \caption{Memory footprint and performance overhead of a GROMACS-DeePMD simulation (red) with respect to a classical MD simulation (blue). Tests are performed on System-1, using the small 1YRF protein, with one MPI process and one GPU.}
    \label{fig_dp_vs_classical}
\end{figure}

\textbf{Classical-MD baseline and overhead.} Before showing scaling results, we discuss the computational and memory overhead introduced by DeePMD relative to standard GROMACS runs, as reported in Fig.~\ref{fig_dp_vs_classical}. For the 1YRF protein on System-1, DeePMD inference reduces throughput by roughly three orders of magnitude compared with classical MD, consistent with other results available in literature~\cite{hu2026gromacsAI_pdp}. At the same time, the measured GPU memory footprint increases from about 0.5 GB for classical MD to about 7 GB for DP-aided MD. The total footprint scales approximately linearly with the size of the NN group, thus, extrapolating this trend to the 1HCI protein (15,000 atoms) yields a requirement above 200~GB, clearly motivating the need for multi-GPU distributed-memory inference even for moderate-sized proteins.

\begin{figure}[!h]
    \centering
    \includegraphics[width=0.9\linewidth]{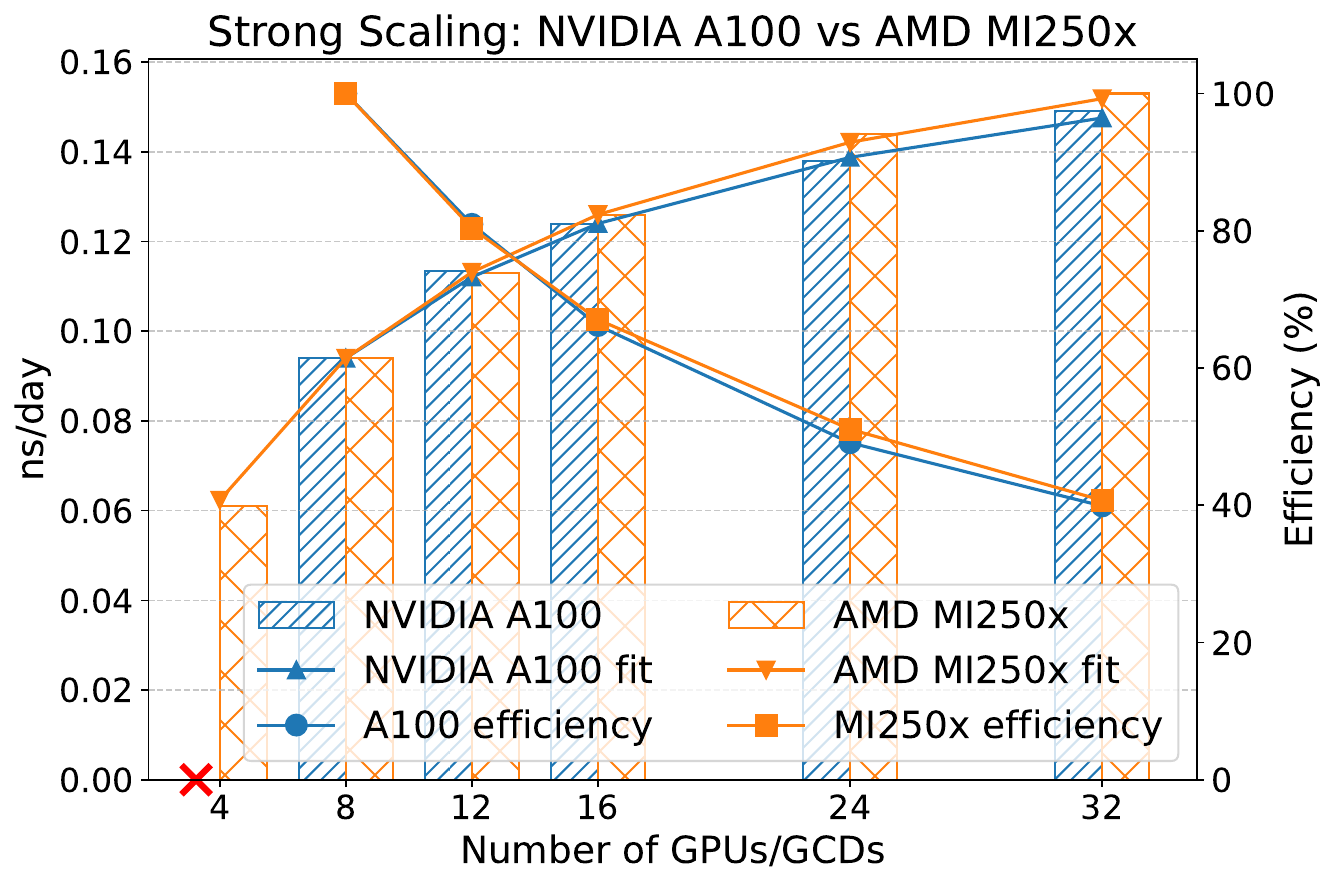}
    \caption{Strong scaling test on NVIDIA A100 and AMD MI250x GPUs for one 1HCI protein. The left y axis shows the measured simulation throughput vs the number of GPUs/GCDs (bars), and the throughput computed according to Eq.~\ref{eq_strong_scaling}. The right y axis reports the scaling efficiency. The simulation cannot be run on four NVIDIA A100 GPUs due to insufficient VRAM.}
    \label{fig_strong_scaling}
\end{figure}

\textbf{Strong scaling} test results for both NVIDIA and AMD systems, scaling from four to 32 devices are reported in Fig.~\ref{fig_strong_scaling}. The PyTorch inference task requires a high amount of memory, and due to the less VRAM available on A100 GPUs compared to MI2050x (40 GB vs 64 GB), the 1HCI protein cannot be simulated using only four A100 devices. For this reason, we use the performance on eight devices as the reference to calculate the scaling efficiency.

NVIDIA and AMD hardware deliver nearly identical performance. Scaling efficiency decreases to 66\% at 16 GPUs and 40\% at 32. The slightly higher throughput observed on AMD at 24–32 devices likely reflects a more favorable inter-node communication pattern. Indeed, System-1 hosts twice as many devices per node as System-2, so the same device count spans half as many nodes on System-1, reducing cross-node MPI traffic.

Physical constraints impose fundamental limits on strong-scaling efficiency. In DD simulations, the number of ghost atoms ($N_{\mathrm{a}}^{\mathrm{ghost}}$) constructed by each subdomain can be assumed independent of the total number of subdomains since it depends primarily on the cutoff radius and not on the subdomain size (if subdomain size and cutoff radius are similar). Thus, the number of atoms processed per MPI rank is approximately $N_{\mathrm{a}}^{\mathrm{tot}}/N_{\mathrm{p}}~+~ N_{\mathrm{a}}^{\mathrm{ghost}}$, where $N_{\mathrm{a}}^{\mathrm{tot}}$ is the total atom count in the system and $N_{\mathrm{p}}$ is the number of processes. Neglecting communication costs, the throughput can be estimated as reported in Eq.~\ref{eq_strong_scaling}:
\begin{equation}
  tr \;=\; \frac{k}{\,N_{\mathrm{a}}^{\mathrm{tot}}/N_{\mathrm{p}} + N_{\mathrm{a}}^{\mathrm{ghost}}\,}
  \;=\; \frac{1}{\,\alpha/N_{\mathrm{p}} + \beta\,},
  \label{eq_strong_scaling}
\end{equation}
where $k$ is a constant taking into account unit conversions and $\alpha = N_{\mathrm{a}}^{\mathrm{tot}}/k$, $\beta = N_{\mathrm{a}}^{\mathrm{ghost}}/k$.
We fit Eq.~\ref{eq_strong_scaling} to measured throughput on System-1 and System-2 for 8 and 16 MPI ranks. The results in Fig.~\ref{fig_strong_scaling} show near perfect agreement between measured and predicted throughput, indicating that the model captures the dominant scaling behavior in this regime. The communicated message is about 28~bytes per NN atom. Even for $10^5$ NN atoms this corresponds to only a few MB per collective, which helps explain why communication remains secondary in the current inference-dominated regime.

\begin{figure}[!h]
    \centering
    \includegraphics[width=0.9\linewidth]{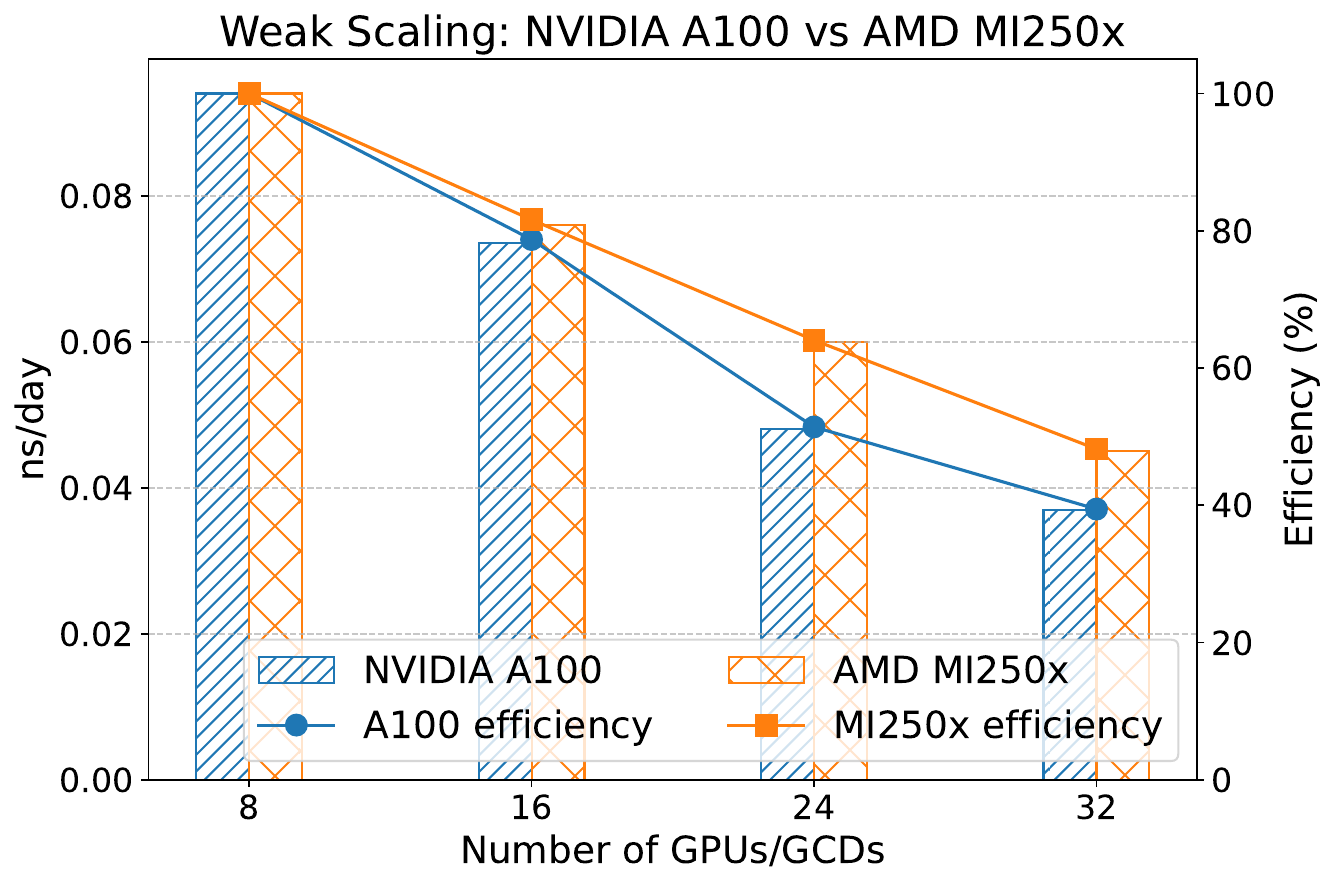}
    \caption{Weak scaling test on NVIDIA A100 and AMD MI250x GPUs for the 1HCI protein. The left y axis shows the measured simulation throughput vs the number of GPUs/GCDs (bars), the right y axis reports the scaling efficiency.}
    \label{fig_weak_scaling}
\end{figure}

\textbf{Weak scaling} results for the two systems are shown in Fig.~\ref{fig_weak_scaling} for simulations from 8 to 32 devices. NVIDIA and AMD GPUs exhibit similar behavior up to 16 devices, maintaining a scaling efficiency of ~80\%. Beyond this point, the MI250x outperforms the A100: at 24 and 32 devices, efficiencies are 64\% and 48\% for MI250x, compared to 51\% and 40\% for A100, respectively. As noted in the strong-scaling tests,
part of the throughput difference likely reflects the fewer compute nodes used on System-1, which reduces the inter-node communication volume. More importantly, the loss of weak-scaling efficiency at 24-32 devices is not explained by communication volume alone. As the replicated system grows, each rank processes its local atoms plus a geometry-dependent ghost population, and small differences in local-plus-ghost counts translate into different inference times. The final collective then exposes this imbalance as synchronization overhead.

\begin{figure*}[h!]
    \centering
    \includegraphics[width=0.95\linewidth]{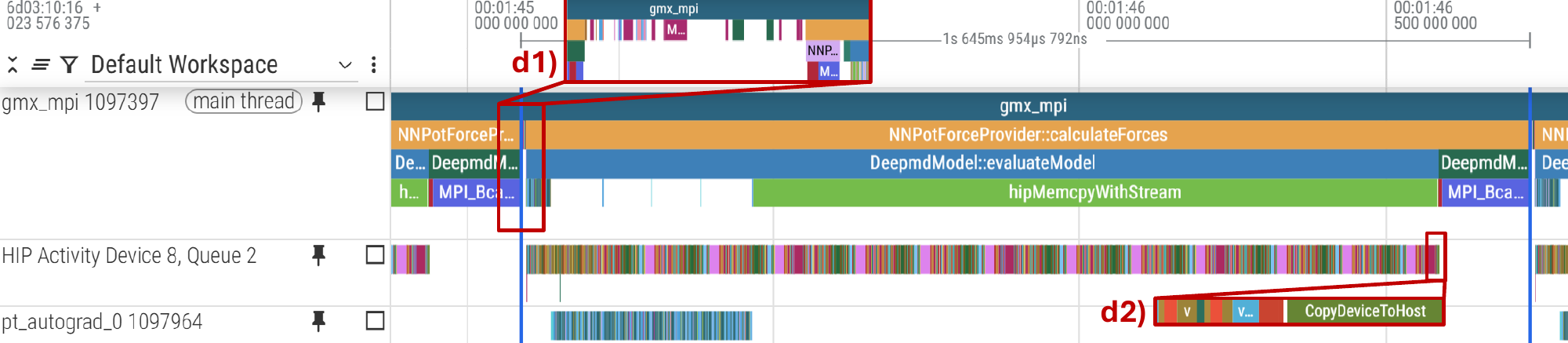}
    \caption{Trace of one MD simulation step obtained with the \texttt{ROCm Systems Profiler} on the system equipped with AMD MI250x GPUs during a simulation on 16 ranks. The trace shows the main CPU thread timeline, the HIP GPU queue, and the PyTorch main CPU thread. The block \texttt{NNPotForceProvider::calculateForces} shows the total time spent in the NNPot module. The standard MD operations are barely visible in the main timeline and are highlighted in the d1) box. \texttt{DeepmdModel::evaluateModel} accounts for the time spent in the inference task. All inference kernels are launched with non-blocking APIs, only the final device-to-host copy uses a blocking call. For this reason the trace shows long time spent in the \texttt{hipMemcpyWithStream} process. In reality, as indicated in the HIP queue timeline, the GPU is executing other kernels. The actual time required by the device-to-host copy is shown in the d2) box.}
    \label{fig_trace_amd}
\end{figure*}

\textbf{Trace \& Load Imbalance.} Fig.~\ref{fig_trace_amd} shows the \texttt{ROCm System Profiler} trace collected on the AMD system while simulating the 1HCI protein with 16 GPUs on 16 MPI processes. For clarity, we display the trace of a single rank over a single MD simulation step. The profile provides detailed insight into CPU and GPU activity, allowing us to identify the main bottlenecks in the code.

In this simulation, one step takes 1.645~s. More than 99\% of the wall time is spent in the NNPot module (in yellow in the picture) which executes DeePMD-related tasks,  while the classical MD operations, highlighted in the d1) box, account for less than 9~ms. Upon entering NNPot, atom positions are exchanged across all the ranks via an MPI collective call (in pink in the d1) box), this operation requires less than 2 ms, thus being negligible. Roughly 90\% of the total time is consumed by the inference task (\texttt{DeepmdModel::evaluateModel}), while the subsequent collective MPI call that distributes forces across all the ranks takes roughly 10\% of the time. This latter MPI call acts as a global synchronization point, which explains its higher cost relative to the initial exchange despite the same message volume being communicated. Indeed, in the presence of load imbalance, all the ranks must wait for the one on which the inference phase takes the longest time before proceeding. The profile therefore indicates that the dominant distributed-memory penalty in this rank-count and atom-count regime is synchronization induced by load imbalance, not raw communication cost.
All HIP kernels used by PyTorch are launched with non-blocking APIs, only the final device-to-host copy is launched with a blocking call. Consequently, the CPU thread appears to spend a long time in \texttt{hipMemcpyWithStream} in the trace, while in reality the GPU is executing kernels in the background, as indicated by the HIP queue timeline.  The actual device-to-host copy time is less than 100~$\mu$s, and it is highlighted in the d2) box.

\section{Discussion \& Conclusion}
\label{sec_discussion_conclusion}
In this work, we extended the functionalities of the classical GROMACS MD package by integrating DeePMD-kit framework, thus enabling the use of DeePMD-family deep potential models in standard MD workflows and thereby achieving near \emph{ab initio} accuracy at substantially lower computational cost than quantum mechanics calculations. The integration leverages the existing GROMACS \texttt{NNPot} module: we extended the interface to support DeePMD-kit APIs and, most importantly, we added full support for GPU-accelerated domain decomposition simulations. As a result, DP-aided runs can be performed on distributed-memory systems, overcoming the memory and compute limitations inherent to single process executions. The native highly optimized GROMACS communication pattern did not readily accommodate DP models. To preserve the core GROMACS design, the DD implementation employed a DD layer decoupled from the one used by the main engine, and two MPI collective operations at each timestep: (i) a collective that shares atomic coordinates to all ranks before DP inference, and (ii) a collective that aggregates and redistributes the DP computed forces across ranks afterward. This approach enabled correct DP coupling without any changes to the core GROMACS engine, preserving all functionalities that users expect.

We extensively assessed the implementation on different systems equipped with NVIDIA A100 and AMD MI250x GPUs, using a DPA-1 model that we trained in-house on a dataset of solvated proteins. For the 1HCI protein (15,668 atoms), strong-scaling tests showed an efficiency of 40\% on both architectures when using 32 devices.  At the same time, weak-scaling tests yielded efficiencies of 51\% on AMD GPUs and 40\% on NVIDIA GPUs at 32 devices. Relative to classical MD, DeePMD-aided runs are roughly three orders of magnitude slower and require more than an order of magnitude more memory for the investigated systems, which is why distributed-memory inference is already necessary for small biomolecular workloads.

In the medium rank-count and small-to-medium protein size ranges, the performance analysis revealed three principal bottlenecks that limit scalability. First, as confirmed by a simple performance model, strong scaling efficiency was not constrained by MPI communication, rather, it was dominated by the number of ghost atoms, which imposes a lower bound on the number of atoms each process must simulate, regardless of the total number of ranks. The ghost atom count is determined by the physical system and the cutoff radius, thus, it cannot be reduced through software engineering solutions.
Second, the dominant memory cost is not the replicated coordinate buffer but the DeePMD inference data structures. Third, detailed profiling with the \texttt{ROCm Systems Profiler} revealed that the time spent in collective communications is negligible compared to the DP inference time. This behavior is expected, since only atoms processed by the DP, not the entire molecular system, are exchanged in these collectives. The dominant bottleneck was therefore imbalance: ranks that perform inference on more atoms take longer, and since the final MPI collective acts as a global synchronization point, the slower ranks determine the overall timestep duration. For substantially larger GPU counts (e.g., $\gtrsim 500$) and/or very large NN groups (several million atoms), the raw communication cost involved in the all-to-all communication may become relevant, and a point-to-point halo exchange (LAMMPS-style) would become the natural design solution.

In this work, we demonstrated that DP-aided production-scale simulations are feasible in GROMACS for local DeePMD models such as DPA-1. Although our analysis indicated that MPI collectives are not the primary bottleneck for the tested scenarios, future work would involve integrating a LAMMPS-style half-shell communication pattern and communication-aware neighbor list into the engine. These changes are key features required for simulations with a large rank count and for seamless support for message-passing DP models such as DPA-2 and DPA-3.

\section*{Acknowledgment}
Financial support from the SeRC Exascale Simulation Software Initiative (SESSI) is gratefully acknowledged.

\ifCLASSOPTIONcaptionsoff
\fi

\bibliographystyle{IEEEtran}
\bibliography{bibtex/IEEEBib}

\end{document}